\documentclass [6pt,a4paper]{article}
\usepackage{bbm}
\usepackage{amsfonts}
\usepackage{mathrsfs}
\usepackage {amssymb}
\usepackage {amsmath}
\usepackage{amsthm}
\usepackage{latexsym}
\def\dse#1{\vskip 0.6cm\noindent
        {\large\bf #1}
        \vskip 0.4cm}
\def\dse#1{\vskip 0.6cm\noindent
        {\large\bf #1}
        \vskip 0.4cm}
\usepackage{lastpage}
\usepackage{epsfig}

\begin{document}
\begin{center}
\textbf{\large{ $\mathbb{Z}_p\mathbb{Z}_p[u]$-additive codes}}\footnote { E-mail
addresses:
luzhenliang1992@sina.cn(Z.lu), zhushixin@hfut.edu.cn(S.Zhu).\\
This research is supported by the National Natural Science
Foundation of China (No.61370089) and the Anhui Provincial Natural Science
Foundation under Grant JZ2015AKZR0229.}
\end{center}

\begin{center}
{ { Zhenliang Lu, \ Shixin Zhu } }
\end{center}

\begin{center}
\textit{\footnotesize Department of Mathematics, Hefei University of
Technology, Hefei 230009, Anhui, P.R.China }
\end{center}

\noindent\textbf{Abstract:} In this paper, we study $\mathbb{Z}_p\mathbb{Z}_p[u]$-additive codes,
where $\emph{p}$ is prime and $u^{2}=0$. In particular, we determine a Gray map from
$ \mathbb{Z}_p\mathbb{Z}_p[u]$ to $\mathbb{Z}_p^{ \alpha+2 \beta}$ and study generator and parity check matrices for these codes. We prove that a Gray map $\Phi$ is a distance preserving map from ($\mathbb{Z}_p\mathbb{Z}_p[u]$,Gray distance) to ($\mathbb{Z}_p^{\alpha+2\beta}$,Hamming distance), it is a weight preserving map as well. Furthermore
we study the structure of $\mathbb{Z}_p\mathbb{Z}_p[u]$-additive cyclic codes.\\

\noindent\emph{\textbf{Keywords}}:additive codes; $\mathbb{Z}_p\mathbb{Z}_p[u]$-additive codes; $\mathbb{Z}_p\mathbb{Z}_p[u]$-additive cyclic codes; Gray map.

\dse{1~~Introduction}
Additive codes with the remarkable paper by Delsarte in 1973[1], he defines additive codes as subgroups of the underlying abelian group in a translation association scheme. In 2006, Borges J. et al. define an extension of the usual Gray map, the new Gray map is an isometry which transforms Lee distance in $Z_{2}^{\alpha}\times Z_{4}^{\beta}$ to Hamming distance in
$Z_{2}^{\alpha+2\beta}$[6]. Then, many properties of additive codes are studied. Two kinds of maximum distance separable codes over $Z_{2}Z_{4}$ are studied[7], all MDS $Z_{2}Z_{4}$-additive codes are zero or one error-correcting codes with the exception of the trivial repetition codes containing two codewords. Cyclic additive codes are also studied[8][15]. Recently, $Z_{2}Z_{4}$-additive codes were generalized to $Z_{2}Z_{2^{s}}$-additive codes by Aydogdu and Siap[9]. And next $Z_{p^{r}}Z_{p^{s}}$-additive codes are studied by Aydogdu and Siap[4]. In [4], the paper given the standard generator matrices and dual matrices of the form over $Z_{p^{r}}Z_{p^{s}}$-additive codes.

Later, in [3], a generalization towards another direction that have a good algebraic structure and provide good binary codes is presented, a new class of additive codes which is referred to as $Z_{2}Z_{2}[u]$-additive codes is introduced. About the application of additive codes to steganography is proposed[10] and lt's also helped to study quantum code. Now, quantum additive code is a new research direction. Many articles and research has been done on quantum additive codes. In this paper, we extend the $Z_{2}Z_{2}[u]$-additive codes to codes over $\mathbb{Z}_p\mathbb{Z}_p[u]$,where  $\emph{p}$ is prime and $u^{2}=0$. Corresponding, we given a more simplify standard generator matrices and dual matrices of the form. At the same time, we define a Gray map $\Phi$. We prove that a Gray map $\Phi$ is a distance preserving map from ($\mathbb{Z}_p\mathbb{Z}_p[u]$,Gray distance) to ($\mathbb{Z}_p^{\alpha+2\beta}$,Hamming distance), it is a weight preserving map as well. At the end of the paper, we study the structure of $\mathbb{Z}_p\mathbb{Z}_p[u]$-additive cyclic codes.\\

\dse{2~~Preliminaries}

 Let \textbf{}$\mathbb{Z}_{p}$ be a finite filed with
$p$ elements, where $p$ is an odd prime. Let $R$ be the commutative
ring $\mathbb{Z}_p+u\mathbb{Z}_p=\{a+ub\mid a,b\in \mathbb{Z}_p\}$
where $u^2=0$. A linear code $C$ over R containing some nonzero codewords
is permutation equivalent to a code with a generator matrix of the form
$$ G= \begin {pmatrix} I_{k_0} & A & B\\0 & uI_{k_{1}} & uD\end{pmatrix},$$
where $A,D$ are $p$-ary matrices, $B$ is $\mathbb{Z}_p+u\mathbb{Z}_p$-matrices, $I_{k_0}$and $I_{k_{1}}$denote the $k_{0}\times k_{0}$ and
$k_{1}\times k_{1}$ identity matrices, and $C$ contains $p^{2k_{0}+k_{1}}$ codewords[2].

We define a Gray map $\psi$ from $R$ to ${Z}_{p}^{2}$ in the following way.
\begin{align*}
\psi:R&\rightarrow {Z}_{p}^{2}\\
 (a+ub)&\rightarrow(b,a+b).
\end{align*}

The set $\mathbb{Z}_p\mathbb{Z}_p[u]$ is defined by
$$\mathbb{Z}_p\mathbb{Z}_p[u]=\{(a,b)|a\in\mathbb{Z}_p \ and \ b\in R\}$$

The set not well defined with respect to the usual multiplication, therefore, to make it well defined
and get some good results, we introduce a new scalar multiplication in the following way:\\
(1)$\forall$ $c_{1}=({a_{0},a_{1},\cdots,a_{\alpha-1},b_{0},b_{1},\cdots,b_{\beta-1}})$,
$c_{2}=({a^{'}_{0},a^{'}_{1},\cdots,a^{'}_{\alpha-1},b^{'}_{0},b^{'}_{1},\cdots,b^{'}_{\beta-1}})\in\mathbb{Z}_p\mathbb{Z}_p[u]$
$$c_{1}c_{2}=({{a_{0}a^{'}_{0},a_{1}a^{'}_{1},\cdots,a_{\alpha-1}a^{'}_{\alpha-1},
b_{0}b^{'}_{0},b_{1}b^{'}_{1},\cdots,b_{\beta-1}b^{'}_{\beta-1}}})$$
(2)$\forall$ $c_{1}=({a_{0},a_{1},\cdots,a_{\alpha-1},b_{0},b_{1},\cdots,b_{\beta-1}})\in\mathbb{Z}_p\mathbb{Z}_p[u]$,
$c=r+qu\in R.$
$$cc_{1}=({ra_{0},ra_{1},\cdots,ra_{\alpha-1},cb_{0},cb_{1},\cdots,cb_{\beta-1}})$$
(3)$\forall$ $c_{1}=({a_{0},a_{1},\cdots,a_{\alpha-1},b_{0},b_{1},\cdots,b_{\beta-1}})\in\mathbb{Z}_p\mathbb{Z}_p[u]$,
$c\in \mathbb{Z}_p.$
$$cc_{1}=({ca_{0},ca_{1},\cdots,ca_{\alpha-1},cb_{0},cb_{1},\cdots,cb_{\beta-1}})$$

\dse{3~~ $\mathbb{Z}_p\mathbb{Z}_p[u]$-additive codes}
In this section, we introduced the definition of the additive codes and the additive dual codes, determine the structure of the generator matrix and dual generator matrix in the standard form of the code.\\

\noindent\textbf{Definition 3.1.}A linear code C is called a $\mathbb{Z}_p\mathbb{Z}_p[u]$ additive code if it is
a $\mathbb{Z}_p+\mathbb{Z}_p[u]$ submodule of $\mathbb{Z}_p^{\alpha}\times\mathbb{Z}_p[u]^{\beta}$ with respect to the scalar multiplication defined in (1),(2),(3). Then the $p$-ary image $\Phi(C)=\textbf{C}$ is called $\mathbb{Z}_p\mathbb{Z}_p[u]$ linear code of
length $n=\alpha+2\beta$ where $\Phi$ is a map from $\mathbb{Z}_p^{\alpha}\times\mathbb{Z}_p[u]^{\beta}$ to $\mathbb{Z}_p^{n}$ defined as\\
$$\Phi(a,b)=({a_{0},a_{1},\cdots,a_{\alpha-1},\psi(b_{0}),\psi(b_{1}),\cdots,\psi(b_{\beta-1})})$$
for all  $ a=({a_{0},a_{1},\cdots,a_{\alpha-1}})\in \mathbb{Z}_p^{\alpha} $,
$ b=(b_{0},b_{1},\cdots,b_{\beta-1})\in \mathbb{Z}_p[u]^{\beta}$.\\

\noindent\textbf{Theorem 3.2.} Let $C$ be a $\mathbb{Z}_p\mathbb{Z}_p[u]$-additive code of type $(p;\alpha,\beta;k_{0},k_{1})$.
Then $C$ is permutation equivalent to a $\mathbb{Z}_p\mathbb{Z}_p[u]$ additive code with the standard form matrix
\begin{align}
G= \begin {pmatrix} I_{k_0} & A & B\\0 & uI_{k_{1}} & uD\end{pmatrix},
\end{align}
where $A,B,D$ are $R$-matrices,$I_{k_0}$and $I_{k_{1}}$denote the $k_{0}\times k_{0}$ and
$k_{1}\times k_{1}$ identity matrices.\\\\%,$C$ contains $p^{2k_{0}+k_{1}}$ codewords
$ Proof $ ~~Since the $\mathbb{Z}_p\mathbb{Z}_p[u]$ additive codes front part is $\mathbb{Z}_p^{\alpha}$,so the $\mathbb{Z}_p\mathbb{Z}_p[u]$ additive codes can be generated by a matrix as follow:
$$ \begin {pmatrix} I_{k_0} & S_{1}\end{pmatrix},$$
where S are $Z_{P}$-matrix.

Likewise, the $\mathbb{Z}_p\mathbb{Z}_p[u]$ additive codes after part is $\mathbb{Z}_p+u\mathbb{Z}_p$,
so the $\mathbb{Z}_p\mathbb{Z}_p[u]$ additive codes can be generated by a matrix as follow:
$$ \begin {pmatrix} S_{2} & I_{k_1} & A_{1} & A_{2}\\S_{3} & 0 & uI_{k_{2}} & uA_{3}\end{pmatrix},$$
where $S_{2},S_{3},A_{1},A_{2},A_{3}$ are $Z_{P}$-matrices.$I_{k_1},I_{k_2}$ is identity matrices.

According to generator matrices theorem,we know the matrices
$$ \begin {pmatrix}I_{k_0} & S_{11} & S_{12} & S_{13} \\ S_{2} & I_{k_1} & A_{1} & A_{2}\\S_{3} & 0 & uI_{k_{2}} & uA_{3}\end{pmatrix},$$
is also generate the additive codes,where $ S_{1}$=$ S_{11}+ S_{12}+ S_{13}$.

Next by applying necessary row and column oprations to the above matrix,we obtain
$$ \begin {pmatrix}I_{k_0} & 0 & S_{12}^{'} & S_{13}^{'} \\ 0 & I_{k_{11}} & A_{1}^{'} & A_{2}^{'}\\0 & 0 & uI_{k_{22}} & uA_{3}^{'}\end{pmatrix},$$

Let $k_{0}^{'}$=$k_{1}$+$k_{11}$,we can obtain the matrices
$$ G= \begin {pmatrix} I_{k_0^{'}} & A & B\\0 & uI_{k_{1}} & uD\end{pmatrix},$$
Finally,Let $k_0^{'}=k_0$,we reach to the claimed form.\qed\\

The inner product for the vectors $v,w\in $$\mathbb{Z}_p\mathbb{Z}_p[u]$ is defined by
\[v\cdot w=u ( \sum_{i=1}^{\alpha}v_{i}w_{i})+\sum_{j=\alpha+1}^{\alpha+\beta}v_{j}w_{j}\in {Z}_p+u{Z}_p\]
\noindent\textbf{Definition 3.3.}Let C be a $\mathbb{Z}_p\mathbb{Z}_p[u]$-additive code,The additive dual code of $C$,denote by $C^{\perp}$, and\\
\[C^{\perp}=\{w\in\mathbb{Z}_p^{\alpha}\times\mathbb{Z}_p[u]^{\beta}\mid v\cdot w=0~ for~all~v\in C\}.\]

\noindent\textbf{Theorem 3.4.}Let $C$ be a $\mathbb{Z}_p\mathbb{Z}_p[u]$ additive code of type $(p;\alpha,\beta;k_{0},k_{1})$ with
the standard form matrix defined in Equation (1),Then the generator matrix for the additive dual code $ C^{\perp}$ is given by
\begin{align}
H= \begin {pmatrix} -B^{t}+D^{t}A^{t} & -D^{t} & I_{n-k_{0}-k_{1}}\\uA^{t} & -uI_{k_{1}} & 0\end{pmatrix},
\end{align}
$ Proof $~~Denote the code with generator matrix (2) by $C^{'}$. Since $ HG^{'}=0$, clearly $C^{'}\in C^{\perp}$.\\
Let $c=(c_{1},c_{2},\cdots,c_{n})\in C^{\perp}$. After adding a linear combination of the first $n-k_{0}-k_{1}$
row of (2) to c, we obtain a codeword is of the form
\[c^{'}=(c_{1},c_{2},\cdots,c_{k_{0}},c_{k_{0}+1},\cdots,c_{k_{0}+k_{1}},0,\cdots,0)\in C^{\perp}\]
Since $c^{'}$ is orthogonal to the last $k_{1}$ rows of (1),so we can adding a certain linear combination of the last $k_{1}$ row of (2) to $c^{'}$. Similar, we obtain a codeword is of the form
\[c^{''}=(c_{1},c_{2},\cdots,c_{k_{0}},0,\cdots,0)\in C^{\perp}\]
Since $c^{''}$ is orthogonal to the first $k_{0}$ rows of (1), so we can obtain $c_{1} = c_{2} = \cdots =c_{k}=0$.
so $c \in C^{'}$, $C^{\perp}  \in  C^{'}$.
Therefore H is the generator matrix of the additive dual code $C^{\perp}.$\qed\\

\noindent\textbf{Example 3.5.} Let $C$ be a $\mathbb{Z}_p\mathbb{Z}_p[u]$-additive code of type $(3;1,4;2,2)$ with the standard form generator matrix:
\begin{align}
G= \begin {pmatrix} 1 & 0 & 0 & 1 & 1\\0 & 1 & 0 & 2 & 0\\ 0 & 0 & u & 0 & 2u\\ 0 & 0 & 0 & u & 0\end{pmatrix}
\end{align}
Then,the parity-check matrix of $C$ as given:
\begin{align}
H= \begin {pmatrix} 2 & 0 & 1 & 0 & 1\\0 & 0 & 2u & 0 & 0\\ u & 2u & 0 & 2u & 0\end{pmatrix}
\end{align}
And it's clear that $C^{\perp}$ is of  type $(3;1.4;1,2)$.\\

Notice that the number of codewords cannot given by the additive code of type.

\dse{4~~The gray map}

In this part of the paper, we study the MacWilliams identity for $\mathbb{Z}_p\mathbb{Z}_p[u]$-additive code,
the results is similar to $p=2$ [3], and a Gray map $\Phi$ is given, we found the Gray map $\Phi$ is a distance preserving map from ($\mathbb{Z}_p\mathbb{Z}_p[u]$,Gray distance) to ($\mathbb{Z}_p^{\alpha+2\beta}$,Hamming distance), and it is also a weight preserving map.\\

In the Preliminaries, we also define a Gray map $\psi$ from $R$ to ${Z}_{p}^{2}$ in the following way.
\begin{align*}
\psi:R&\rightarrow {Z}_{p}^{2}\\
 (a+ub)&\rightarrow(b,a+b).
\end{align*}

At the same time, in definition 3.1., we given a map $\Phi$, it is from $\mathbb{Z}_p^{\alpha}\times\mathbb{Z}_p[u]^{\beta}$ to $\mathbb{Z}_p^{n}$ defined as\\
$$\Phi(a,b)=({a_{0},a_{1},\cdots,a_{\alpha-1},\psi(b_{0}),\psi(b_{1}),\cdots,\psi(b_{\beta-1})})$$
for all  $ a=({a_{0},a_{1},\cdots,a_{\alpha-1}})\in \mathbb{Z}_p^{\alpha} $,
$ b=(b_{0},b_{1},\cdots,b_{\beta-1})\in \mathbb{Z}_p[u]^{\beta}$.\\

Let C be an additive code and assume $n=\alpha+2\beta$, the weight enumerator of an additive code $C$
 is defined by
 \[W(x,y)=\sum_{c\in C}x^{n-w(c)}y^{w(c)}.\]\\

\noindent\textbf{Theorem 4.1.} Let $C$ be a $\mathbb{Z}_p\mathbb{Z}_p[u]$-additive code, and $C^{\perp}$
be its dual code, then their weight enumerators $W_{G}(x,y)$ and $W_{G^{\perp}}(x,y)$ are connected by the MacWilliams identity:
\[W_{G^{\perp}}(x,y)=\frac{1}{|c|}W_{G}(X+(q-1)Y,X-Y)\]
$ Proof $~~Similar to the proof of [3,theorem 3.3].\qed\\

Let $F_{p}^{*}$ is a multiplication group with
nonzero elements, where $p$ is an odd prime. Next we definition a Gray weight $W_{G}(c)$ for $c=(c_{1},c_{2},\cdots,c_{n})$
in the following way:
\[W_{G}(c)=\sum_{i=0}^{n-1}W_{G}(c_{i})\]
where
\begin{eqnarray*}
 W_{G}(c_{i})=
\left\{ {{\begin{array}{ll}
 {0}, & {\textrm{if}\mbox{ } c_{i} = 0 },\\
 {2}, & {\textrm{if}\mbox{ } c_{i}=a+u(p-b),a,b\in F_{p}^{*}~and ~a\neq b}, \\
 {1,} & {\textrm{}\mbox{ } others}. \\
\end{array} }} \right .
\end{eqnarray*}
This gray weight function defines also a gray distance function
\[d_{G}(x,y)=W_{G}(x-y)\]
The Hamming weight of a weight of $n$-tuples is the number of its nonzero entries. The Hamming distance between
two $n$-tuples is defined as the Hamming weight of their difference.
Denote the Hamming weight of a weight of a $p$-ary vector $x$ by $W_{H}(x)$ and the Hamming distance between two $p$-ary
vectors $x$ and $y$ of the same length by $d_{H}(x,y)$, and we have $W_{H}(x-y)=d_{H}(x,y)$. \\

Since $\forall$ $c=(c_{1},c_{2},\cdots,c_{n})\in \mathbb{Z}_p\mathbb{Z}_p[u]$. We have
\begin{eqnarray*}
 W_{H}(\Phi(c_{i}))=
\left\{ {{\begin{array}{ll}
 {0}, & {\textrm{if}\mbox{ } c_{i} = 0 },\\
 {2}, & {\textrm{if}\mbox{ } c_{i}=a+u(p-b),a,b\in F_{p}^{*}~and ~a\neq b}, \\
 {1,} & {\textrm{}\mbox{ } others}. \\
\end{array} }} \right .
\end{eqnarray*}

Clearly, $W_{G}(c_{i})=W_{H}(\Phi(c_{i}))$~$\forall$ $c_{i}\in \mathbb{Z}_p,~~i\in(1,2,\cdots,n).$\qed\\

\noindent\textbf{Theorem 4.2.} The Gray map $\Phi$ is a weight preserving map from
\[(\mathbb{Z}_p^{\alpha}\mathbb{Z}_p[u]^{\beta},Gray~~ weight)~~~to~~~(\mathbb{Z}_p^{\alpha+2\beta},Hamming ~~weight)\]
i.e.
\begin{align}
W_{G}(c)=W_{H}(\Phi(c)~~~for~ \forall ~c~\in \mathbb{Z}_p\mathbb{Z}_p[u].
\end{align}
and  $\Phi$ is a distance preserving map from
\[(\mathbb{Z}_p^{\alpha}\mathbb{Z}_p[u]^{\beta},Gray~~ distance)~~~to~~~(\mathbb{Z}_p^{\alpha+2\beta},Hamming ~~distance)\]
i.e.
\begin{align}
d_{G}(x,y)=d_{H}(\Phi(x),\Phi(y))~~~for~ \forall ~x,y~\in \mathbb{Z}_p\mathbb{Z}_p[u].
\end{align}
$ Proof $~~Let $\forall$ $c=(c_{1},c_{2},\cdots,c_{\alpha},c_{\alpha+1},\cdots,c_{\alpha+\beta})\in \mathbb{Z}_p^{\alpha}\mathbb{Z}_p[u]^{\beta}$,
where $c_{i}\in \mathbb{Z}_p^{\alpha},i=1,2,\cdots,\alpha.$
$c_{\alpha+i}=r_{i}+uq_{i}\in \mathbb{Z}_p[u]^{\beta},i=1,2,\cdots,\beta.$
by the grap map $\Phi$ we obtain:
\begin{align*}
\Phi(c)&=(c_{1},c_{2},\cdots,\psi(c_{\alpha}),\psi(c_{\alpha+1}),\cdots,\psi(c_{\alpha+\beta}))\\
&=(c_{1},c_{2},\cdots,c_{\alpha},q_{1},q_{2},\ldots,q_{\beta},q_{1}+r_{1},q_{2}+r_{2},\cdots,q_{\beta}+r_{\beta})
\end{align*}
\begin{align*}
W_{H}(\Phi(c))&=W_{H}(c_{1},c_{2},\cdots,c_{\alpha},q_{1},q_{2},\ldots,q_{\beta},q_{1}+r_{1},q_{2}+r_{2},\cdots,q_{\beta}+r_{\beta})\\
&=\sum_{i=1}^{\alpha}W_{H}(c_{i})+\sum_{i=1}^{\beta}W_{H}(q_{i},q_{i}+r_{i})\\
&=\sum_{i=1}^{\alpha}W_{H}(c_{i})+\sum_{i=1}^{\beta}W_{H}(\psi(c_{\alpha+i}))\\
&=\sum_{i=1}^{\alpha}W_{G}(c_{i})+\sum_{i=1}^{\beta}W_{G}(c_{\alpha+i})\\
&=\sum_{i=1}^{\alpha+\beta}W_{G}(c_{i})=W_{G}(c)
\end{align*}
Therefore we have (5).
Similarly,we also can deduce (6),the proof is omitted.\qed\\

\dse{5~~The structure of $\mathbb{Z}_p\mathbb{Z}_p[u]$-additive cyclic code}

In this part of the paper, we introduce the definition of a additive cyclic code and some algebraic structure.
A code $C$ is cyclic if and only if its polynomial representation is an ideal.
Let $R_{\alpha,\beta}[x]=\frac{Z_{p}[x]}{<x^{\alpha}-1>}\times\frac{R[x]}{<x^{\beta}-1>}$.
\\

\noindent\textbf{Definition 5.1.}A additive code $C$ is called a $\mathbb{Z}_p\mathbb{Z}_p[u]$-additive cyclic code if any cyclic shift of a codeword is also a code.
i.e.,
\[ (a_{0},a_{1},\cdots,a_{\alpha-1},b_{0},b_{1},\cdots,b_{\beta-1})\in C
\Rightarrow  (a_{\alpha-1},a_{0},\cdots,a_{\alpha-2},b_{\beta-1},b_{0},\cdots,b_{\beta-2})\in C .\]

\noindent\textbf{Theorem 5.2.}~~If $C$ be any $\mathbb{Z}_p\mathbb{Z}_p[u]$-additive cyclic code, then $C^{\perp}$ is also cyclic.\\
$ Proof $~~~Let $C$ be any $\mathbb{Z}_p^{\alpha}\mathbb{Z}_p[u]^{\beta}$-additive cyclic code. Suppose
$ v=(a_{0},a_{1},\cdots,a_{\alpha-1},b_{0},b_{1},\cdots,b_{\beta-1})\in C^{\perp} $ , for any codeword
$ w=(d_{0},d_{1},\cdots,d_{\alpha-1},e_{0},e_{1},\cdots,e_{\beta-1})\in C $ £¬we have
\[v\cdot w=u ( \sum_{i=0}^{\alpha-1}a_{i}d_{i})+\sum_{j=0}^{\beta-1}b_{j}e_{j}=0\]
Let $S$ is a cyclic shift, and $j=lcm(\alpha,\beta)$. Then we have $S(v)=(a_{\alpha-1},a_{0},\cdots,a_{\alpha-2},b_{\beta-1},b_{0},\cdots,b_{\beta-2})$~~\\
and~~$S^{j}(w)=w$ for any $w\in C$. Since $C$ be any $\mathbb{Z}_p^{\alpha}\mathbb{Z}_p[u]^{\beta}$-additive cyclic code,
So we have \\
\[S^{j-1}(w)=(d_{1},d_{2},\cdots,d_{\alpha-1},d_{0},e_{1},e_{2},\cdots,e_{\beta-1},e_{0})\in C\]
Hence
\begin{align*}
0=v\cdot S^{j-1}(w)=&u(a_{0}d_{1}+a_{1}d_{2}+\cdots+a_{\alpha-2}d_{\alpha-1}+a_{\alpha-1}d_{0})\\
&+(b_{0}e_{1}+b_{1}e_{2}+\cdots+b_{\beta-2}e_{\beta-1}+b_{\beta-1}e_{0})\\
=&u(a_{\alpha-1}d_{0}+a_{0}d_{1}+\cdots+a_{\alpha-2}d_{\alpha-1})\\
&+(b_{\beta-1}e_{0}+b_{1}e_{2}+\cdots+b_{\beta-2}e_{\beta-2})\\
=&S(v)\cdot w
\end{align*}
Therefore,we have $ S(v)\in C^{\perp} $,so $ C^{\perp} $ is a cyclic code.\qed\\

Let $C$ be a $\mathbb{Z}_p\mathbb{Z}_p[u]$-additive cyclic code, for any codeword
$ c=(a_{0},a_{1},\cdots,a_{\alpha-1},b_{0},b_{1},\cdots,b_{\beta-1})\in C $ can be representation with a polynomial,i.e.,
\[ c(x)=(a_{0}+a_{1}x+\cdots+a_{\alpha-1}x^{\alpha-1},b_{0}+b_{1}x+\cdots+b_{\beta-1}x^{\beta-1})=(a(x),b(x))\in R_{\alpha,\beta}[x]. \]
Similarly. In preliminaries, we introduce a new scalar multiplication. Now, we have the following multiplication:\\
(1)$\forall$ $ c_{1}(x)=(a_{1}(x),b_{1}(x)),c_{2}(x)=(a_{2}(x),b_{2}(x))\in R_{\alpha,\beta}[x] ,$
$$c_{1}(x)c_{2}(x)=(a_{1}(x)a_{2}(x),b_{1}(x)b_{2}(x))$$
(2)$\forall$ $ c_{1}(x)=(a_{1}(x),b_{1}(x))\in R_{\alpha,\beta}[x],$
$c_{2}(x)=r(x)+uq(x)\in R[x],$where $ r(x),q(x)\in {Z}_p[x],$
$$c_{1}(x)c_{2}(x)=(a_{1}(x)r(x),b_{1}(x)c_{2}(x))$$
(3)$\forall$ $ c_{1}(x)=(a_{1}(x),b_{1}(x))\in R_{\alpha,\beta}[x],$ $c_{2}(x)\in {Z}_p[x],$
$$c_{1}(x)c_{2}(x)=(a_{1}(x)c_{2}(x),b_{1}(x)c_{2}(x))$$
Clearly, definition 5.1 is equivalent to
\begin{align*}
c(x)&=(a_{0}+a_{1}x+\cdots+a_{\alpha-1}x^{\alpha-1},b_{0}+b_{1}x+\cdots+b_{\beta-1}x^{\beta-1})\in R_{\alpha,\beta}[x].\\
\Longrightarrow xc(x)&=(a_{\alpha-1}+a_{0}x+\cdots+a_{\alpha-2}x^{\alpha-1},b_{\beta-1}+b_{0}x+\cdots+b_{\beta-2}x^{\beta-1})\in R_{\alpha,\beta}[x].
\end{align*}
Now, we define the homomorphism mapping:
\begin{align*}
&\Psi:R_{\alpha,\beta}[x]\longrightarrow R[x]\\
&\Psi(c(x))=\Psi(a(x),b(x))=b(x)
\end{align*}
It is clear that $Image(\Psi)$ is an ideal in the ring ~$\frac{R[x]}{<x^{\beta}-1>}$ and $ker(\Psi)$ is also an ideal over $Z_{p}[x]$. And note that
\[Image(\Psi)=\{b(x)\in R[x]:(a(x),b(x))\in R_{\alpha,\beta}[x]\}\]
\[ker(\Psi)=\{(a(x),0)\in R_{\alpha,\beta}[x]:a(x)\in \frac{Z_{p}[x]}{x^{\alpha}-1})\}\]
By using the characterization in [14], we have \[Image(\Psi)=<g(x)+up(x),uq(x)>\]
where $g(x),p(x),q(x)\in \frac{R[x]}{<x^{\beta}-1>}$,~$~q(x)\mid g(x)\mid (x^{\beta}-1)$ and $q(x)\mid p(x)\frac{x^{\beta}-1}{g(x)}$.\\
Similarly,\[ker(\Psi)=<(f(x),0)>\]
where $ f(x)\in \frac{Z_{p}[x]}{x^{\alpha}-1}$ and $ f(x)\mid (x^{\alpha}-1) $.\\

According to the homomorphism map theorem we have:
\[C/ ker(\Psi)\cong <g(x)+up(x),uq(x)>.\]
Hence, we have
\[(h(x),(g(x)+up(x),uq(x))\in C\]
where $\Psi(h(x),(g(x)+up(x),uq(x)))=(g(x)+up(x),uq(x))$.\\

By these discussion, it is easy to see that any $\mathbb{Z}_p\mathbb{Z}_p[u]$-additive cyclic code can be generated by two elements of the form $(h(x),(g(x)+up(x),uq(x)))$ and $(f(x),0)$.\\

\noindent\textbf{Corollary 5.3.}Let $C$ be a $\mathbb{Z}_p\mathbb{Z}_p[u]$-additive cyclic code. Then $C$ is an ideal in $R_{\alpha,\beta}[x]$ which can be generated by
\[C=((f(x),0),(h(x),(g(x)+up(x),uq(x)))).\]
where~$~q(x)\mid g(x)\mid (x^{\beta}-1)$, $q(x)\mid p(x)\frac{x^{\beta}-1}{g(x)}$ .\qed\\

\noindent\textbf{Corollary 5.4.} Let $C=((f(x),0),(h(x),(g(x)+up(x),uq(x))))$ is a $\mathbb{Z}_p\mathbb{Z}_p[u]$-additive cyclic code,
then we may assume that  $f(x)\mid h(x)\frac{x^{\beta}-1}{l(x)}$ .where $l(x)=lcm(p(x),q(x))$.\\

$ Proof $~~(1)Since $\Psi(\frac{x^{\beta}-1}{l(x)}(h(x),(g(x)+up(x),uq(x))))=\Psi((\frac{x^{\beta}-1}{l(x)}*h(x),0))=0.$\\
Hence $(\frac{x^{\beta}-1}{l(x)}*h(x),0)\in ker(\Psi)\subseteq C$ and $f(x)\mid h(x)\frac{x^{\beta}-1}{l(x)}$.\qed\\

As a consequence to this corollary, we classify the structure of the additive cyclic code into three categories by the following theorem.\\

\noindent\textbf{Theorem 5.5.} Let $C$ be a $\mathbb{Z}_p\mathbb{Z}_p[u]$-additive cyclic code.Then $C$ can be identified as following:\\
(1)$C=((f(x),0)$, where $ f(x)\in \frac{Z_{p}[x]}{x^{\alpha}-1}$ .\\
(2)$C=(h(x),(g(x)+up(x),uq(x)))$, where $~q(x)\mid g(x)\mid (x^{\beta}-1)$ and $(x^{r}-1)\mid p(x)\frac{x^{\beta}-1}{g(x)}$.\\
(3)$C=((f(x),0),(h(x),(g(x)+up(x),uq(x))))$,where ~$~q(x)\mid g(x)\mid (x^{\beta}-1)$, $q(x)\mid p(x)\frac{x^{\beta}-1}{g(x)}$, $f(x)\mid h(x)\frac{x^{\beta}-1}{l(x)}$ and $l(x)=lcm(p(x),q(x))$.\qed\\

\noindent\textbf{Corollary 5.6.}Let $C$ be any $\mathbb{Z}_p\mathbb{Z}_p[u]$-additive cyclic code. Then $\Phi(C)$ is an cyclic code of length $\alpha+2\beta$ over $Z_{p}$.\\

$ Proof $~~Let $S$ is a cyclic shift. Since $C$ be any $\mathbb{Z}_p\mathbb{Z}_p[u]$-additive cyclic code. For any codeword
$$ c=(a_{0},a_{1},\cdots,a_{\alpha-1},b_{0},b_{1},\cdots,b_{\beta-1})\in C $$
where $b_{i}=r_{i}+uq_{i},i\in \{0,1,2,\cdots,\beta-1\}, a_{i},r_{i},q_{i}\in Z_{p}$.\\
We have $$ S(c)=(a_{\alpha-1},a_{0},\cdots,a_{\alpha-2},b_{\beta-1},b_{0},\cdots,b_{\beta-2})\in C $$
Then
\begin{align*}
\Phi(S(c))=(&a_{\alpha-1},a_{0},\cdots,a_{\alpha-2},q_{\beta-1},q_{0},\cdots,\\
&q_{\beta-2},q_{\beta-1}+r_{\beta-1},q_{0}+r_{0},\cdots,q_{\beta-2}+r_{\beta-2})\in\Phi(C)
\end{align*}
Then by the Gray map we have:\\ $$\Phi(c)=(a_{0},a_{1},\cdots,a_{\alpha-1},q_{0},q_{1},\cdots,q_{\beta-1},q_{0}+r_{0},q_{1}+r_{1},\cdots,q_{\beta-1}+r_{\beta-1})\in\Phi(C).$$
Hence
\begin{align*}
S(\Phi(c))=(&a_{\alpha-1},a_{0},\cdots,a_{\alpha-2},q_{\beta-1},q_{0},\cdots,q_{\beta-2},\\
&q_{\beta-1}+r_{\beta-1},q_{0}+r_{0},\cdots,q_{\beta-2}+r_{\beta-2})=\Phi(S(c))\in\Phi(C).
\end{align*}
This proves that $\Phi(C)$ is an cyclic code of length $\alpha+2\beta$ over $Z_{p}$.\qed\\

\dse{6~~Conclusion}
In this paper, we studied $\mathbb{Z}_p\mathbb{Z}_p[u]$-additive codes some property, including generator and parity check matrices for the codes. We fund the Gray map $\Phi$ is a distance preserving map and weight preserving map as well. At the end of the paper,we introduce the structure of $\mathbb{Z}_p\mathbb{Z}_p[u]$-additive cyclic code.
The studies makes this family of codes become widespread. we hope this family of codes haven more studies, such as constacyclic codes, depth distribution and other place. Due to this family of codes is newly introduced, some similar problems are still open here. \\

\end{document}